\documentclass[twocolumn,floats,floatfix,aps,pra]{revtex4-1}
\usepackage{amsfonts,amssymb,amsmath}
\usepackage{color,calc}
\usepackage[dvips]{graphicx}
\usepackage{bm}

\def\be{ \begin{equation} }
\def\ee{ \end{equation} }
\def\bea{ \begin{eqnarray} }
\def\eea{ \end{eqnarray} }
\def\bse{ \begin{subequations} }
\def\ese{ \end{subequations} }
\def\ba{ \begin{array} }
\def\ea{ \end{array} }

\def\to{\rightarrow}
\def\fromto{\leftrightarrow}

\def\H{\mathbf{H}}
\def\W{\mathbf{W}}
\def\c{\mathbf{c}}

\newcommand{\ket}[1]{\vert #1\rangle}
\newcommand{\bra}[1]{\langle #1\vert}

\def\etal{\textit{et al.}}

\def\NAC{\chi}

\def\Q{\Omega_q}
\def\Q{Q}
\def\Q{\Omega}

\begin{document}

%%%%%%%%%%%%%%%%%%%%%%%%%%%%%%%%%%%%%%%%%%%%%%%%%%%%%%%%%%%%%%%%%%%%%%%%%%%%%%%%%%%%%%%%%%%%%%%%%%%%%%%%%%%%%%%%%%%%%%%%%%%%%%%%%%%%%%%%%
%%% TITLE PAGE %%%

%\title{High-fidelity multistate STIRAP by ``shortcuts to adiabaticity''}
\title{High-fidelity multistate STIRAP assisted by shortcut fields}

\author{Nikolay V. Vitanov}

\affiliation{Department of Physics, St Kliment Ohridski University of Sofia, 5 James Bourchier blvd, 1164 Sofia, Bulgaria}

\date{\today }

\begin{abstract}
Multistate stimulated Raman adiabatic passage (STIRAP) is a process which allows for adiabatic population transfer between the two ends of a chainwise-connected quantum system.
The process requires large temporal areas of the driving pulsed fields (pump and Stokes) in order to suppress the nonadiabatic couplings and thereby to make adiabatic evolution possible.
%The largest system in which multistate STIRAP has been demonstrated contains nine states.
To this end, in the present paper a variation of multistate STIRAP, which accelerates and improves the population transfer, is presented.
In addition to the usual pump and Stokes fields it uses shortcut fields applied between the states, which form the dark state of the system.
The shortcuts cancel the couplings between the dark state and the other adiabatic states thereby resulting (in the ideal case) in a unit transition probability between the two end states of the chain.
Specific examples of five-state systems formed of the magnetic sublevels of the transitions between two degenerate levels with angular momenta  $J_g=2$ and $J_e=1$ or $J_e=2$ are considered in detail, for which the shortcut fields are derived analytically.
The proposed method is simpler than the usual ``shortcuts to adiabaticity'' recipe, which prescribes shortcut fields between all states of the system, while the present proposal uses shortcut fields between the sublevels forming the dark state only.
%Several configurations of the shortcut fields are considered.
The results are of potential interest in applications where high-fidelity quantum control is essential, e.g. quantum information, atom optics, formation of ultracold molecules, cavity QED, etc.
%Although the resulting evolution is nonadiabatic and the term ``shortcuts to adiabaticity'' is misleading the concept is very useful.
\end{abstract}

\maketitle

%\blue

%%%%%%%%%%%%%%%%%%%%%%%%%%%%%%%%%%%%%%%%%%%%%%%%%%%%%%%%%%%%%%%%%
%%%%%%%%%%%%%%%%%%%%%%%%%%%%%%%%%%%%%%%%%%%%%%%%%%%%%%%%%%%%%%%%%
%%%%%%%%%%%%%%%%%%%%%%%%%%%%%%%%%%%%%%%%%%%%%%%%%%%%%%%%%%%%%%%%%%%%%%%%%%%%%%%%%%%%%%%%%%%%%%%%%%%%%%%%%%%%%%%%%%%%%%%%%%%%%%%%%%%%%%%%%
\section{Introduction}\label{Sec:intro}
%%%%%%%%%%%%%%%%%%%%%%%%%%%%%%%%%%%%%%%%%%%%%%%%%%%%%%%%%%%%%%%%%%%%%%%%%%%%%%%%%%%%%%%%%%%%%%%%%%%%%%%%%%%%%%%%%%%%%%%%%%%%%%%%%%%%%%%%%
%

%\textbf{NOTATION!!!!!!!!}
%\textbf{ZEEMAN POSITIVE/NEGATIVE!!!!!!}
%\textbf{dashed/dotted in FIGURES!!!}

Adiabatic evolution of quantum systems is a concept as old as quantum mechanics \cite{Born1928}.
Regardless of its numerous formulations over the years, one feature is common: in the adiabatic limit there are no transitions between the adiabatic states, defined as the instantaneous eigenstates of the Hamiltonian.
If the system begins its evolution in a single adiabatic state and evolves adiabatically then it will remain in this state at all times, with the only change being the possible accumulation of adiabatic and/or geometric phases.
If the Hamiltonian is time-dependent then the adiabatic states will be time-dependent too, which means that their composition of original diabatic (also known as bare or unperturbed) states will change in time.
Popular adiabatic-passage techniques, such as rapid adiabatic passage (RAP) in two-state systems and stimulated Raman adiabatic passage (STIRAP) in three-state systems \cite{Vitanov2001ARPC,Vitanov2017RMP,Bergmann2019}, use adiabatic states that are equal to different diabatic states in the beginning and the end, thereby achieving adiabatic population transfer between different diabatic states.

In RAP, the diabatic energies are made to cross at a certain instant of time by varying the frequency of the driving field (via frequency chirping) or the transition frequency (via electric or magnetic fields).
The coupling between the two states makes the diabatic level crossing show up as an avoided crossing between the adiabatic states.
This level crossing flips the composition of the adiabatic states: away from the crossing each of them is dominated by one diabatic state before the crossing and by the other one after the crossing.

In STIRAP, the population transfer is carried out via the dark state --- a time-dependent eigenstate of the Hamiltonian involving the two end states 1 and 3 of the three-state chain system $1\fromto 2\fromto3$.
If the system is initially in state 1, and if the Stokes pulse driving the transition $2\fromto 3$ between the initially unpopulated states 2 and 3 is applied before, and vanishes before the pump pulse driving the transition $1\fromto 2$ (counterintuitive pulse order), then the dark state is associated with state 1 in the beginning and state 3 in the end.
Therefore, adiabatic evolution during which the system will remain in the dark state, will completely transfer the population from state 1 to state 3.
An added bonus is that the middle state 2, which is subjected to population decay in many physical implementations, is not populated during the process because it is not present in the dark state.

Both RAP and STIRAP have been extended to multistate systems in numerous papers, see Refs.~\cite{Vitanov2001ARPC,Vitanov2017RMP,Bergmann2019} for reviews.
The great advantage of adiabatic passage techniques is the robustness of the population transfer to variations in various experimental parameters, such as the pulse amplitude, duration, frequency, chirp and shape.
An additional and unique advantage of STIRAP is its resilience to decay from the middle state, as mentioned above.
However, adiabatic techniques are slower than the resonant techniques and their efficiency is less than 100\%.
This imperfect efficiency derives from nonadiabatic losses --- unwanted transitions between the population-carrying adiabatic state and the other adiabatic states due to nonadiabatic couplings.
The latter are generated by the time dependence of the Hamiltonian.
%When considered in the adiabatic basis, adiabatic evolution is accompanied by nonadiabatic couplings generated by the time dependence of the Hamiltonian.
The nonadiabatic couplings lead to leaks of population from the populated adiabatic state with an ensuing loss of population transfer efficiency.

To this end, various proposals have been made for suppressing the nonadiabatic couplings and boosting the population transfer efficiency.
In the context of RAP, Gu\'erin \etal~\cite{Guerin2002,Lacour2008} proposed to shape the Hamiltonian elements --- the Rabi frequency $\Omega(t)$ and the detuning $\Delta(t)$ --- in such a manner that the eigenvalues become parallel, i.e. $\Omega(t)^2 + \Delta(t)^2 =$~const.
This idea follows from seminal papers by Dykhne \cite{Dykhne1960} and Davis and Pechukas \cite{Davis1976} who derived the first-order approximation to the probability for nonadiabatic transitions in terms of the (complex-valued) zeroes of the Hamiltonian eigenvalues (named transition points).
Constant eigenvalues mean no transition points and hence no nonadiabatic transitions to the first order.
This idea was successfully extended to STIRAP \cite{Vasilev2009}.

%***************************************************************
\begin{figure}[t]
\includegraphics[width=0.90\columnwidth]{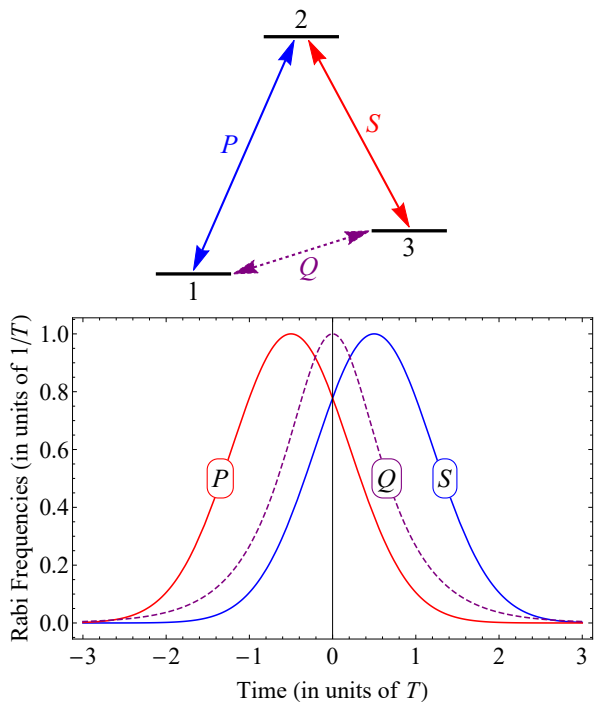}
\caption{
\emph{Top:} Three-state system for STIRAP with the pump (P) and Stokes (S) couplings for the transitions $1\fromto 2$ and $2\fromto 3$ indicated.
The shortcut field $\Q$ drives the transition $1\fromto 3$.
\emph{Bottom:} Pulse shapes of the pump (P), Stokes (S) and shortcut (Q) fields in three-state STIRAP.
}
\label{fig:3ss}
\end{figure}
%***************************************************************

A few years before the pulse-shaping proposals, Unanyan \etal~\cite{Unanyan1997} proposed a rather different idea: apply a third field in STIRAP which directly links states 1 and 3 and exactly matches the nonadiabatic coupling but has the opposite sign, see Fig.~\ref{fig:3ss}.
This third field cancels the nonadiabatic coupling and leads to a perfect population transfer $1 \to 3$.
Later, this approach has been used under different names \cite{Demirplak2003}, with the term ``shortcuts to adiabaticity'' \cite{Chen2010} finally established.
[It is important to point out that the term ``shortcut to adiabaticity'' is misleading because the ensuing evolution, which leads to a perfect population transfer, is actually nonadiabatic.
Indeed, with the third field on the transition $1\fromto 3$ added, the Hamiltonian becomes different and the adiabatic states are different too.
In the adiabatic basis of the new Hamiltonian the evolution is nonadiabatic because there are nonzero couplings and transitions between the adiabatic states.]
Recently, it has been demonstrated experimentally \cite{Versalainen2019} and has been used for chiral resolution \cite{Vitanov2019}.

In the present paper, I use the ideas of Unanyan \etal~\cite{Unanyan1997} for three-state STIRAP and derive shortcut fields which cancel the nonadiabatic couplings in multistate STIRAP and ensure very high population transfer efficiency.
High transfer efficiency is essential in various applications, e.g. in quantum information \cite{Nielsen2000}, or in repeated processes because the single-pass error scales quadratically with the number of processes \cite{Vitanov2018,Vitanov2020}.
I consider the simplest, resonant version of multistate STIRAP, which provides the highest speed of population transfer.
It has been shown  that in the resonant case STIRAP is always possible in systems with an odd number of states \cite{Shore1991,Marte1991,Smith1992,Vitanov1998}, while it is impossible in systems with an even number of states \cite{Band1991,Smith1992,Oreg1992,Vitanov1998}.
In the off-resonance case, STIRAP-like population transfer can take place for both even and odd number of states if the detunings satisfy certain inequalities \cite{Vitanov1998,Vitanov1998epjd}.
Multistate STIRAP has been proposed and experimentally demonstrated for numerous applications, e.g. atomic mirrors and beams splitters in atom optics \cite{Lawall1994,Weitz1994prl,Weitz1994pra,Pillet1993,Valentin1994,Goldner1994,Theuer1998,Featonby1996,Featonby1998,Godun1999,Godun1999jpb}, cavity QED \cite{Parkins1993,Parkins1995}, production of ultracold molecules from ultracold atoms \cite{Winkler2007,Danzl2008,Danzl2010,Ni2008,Lang2008,Aikawa2010,Stellmer2012,Takekoshi2014,Molony2014,Molony2016,Park2015,Guo2016,Rvachov2017,DeMarco2019,Seesselberg2018} atomic clocks \cite{Chalupczak2005}, etc.

Specifically, I consider the five-state systems formed of the three magnetic sublevels $m_g=0,\pm2$ of a ground level with an angular momentum $J_g=2$ coupled to the magnetic sublevels $m_e=\pm1$ of an excited level with $J_e=1$ or $J_e=2$ by left and right circularly polarized laser fields.
STIRAP has already been demonstrated in such systems in the context of atom optics \cite{Theuer1998}.
I propose here to use shortcut fields which compensate the nonadiabatic couplings and ensure very high population transfer efficiency, which is essential in quantum information and atom optics.
The shortcut fields couple only the ground-level sublevels $m_g=0,\pm2$, and hence such couplings can be provided by radiofrequency (rf) fields.

This paper is organized as follows.
Section \ref{Sec:STIRAP} reviews the basic theory of shortcuts and the shortcut to three-state STIRAP.
The standard ``shortcuts to adiabaticity'' for five-state STIRAP in the $J_g=2 \to J_e=1$ and  $J_g=2 \to J_e=2$ systems are derived in Sec.~\ref{Sec:standard}.
Section \ref{Sec:reduced} presents the theory of reduced shortcuts, which allow for easier implementation, with examples in  the $J_g=2 \to J_e=1$ and  $J_g=2 \to J_e=2$ systems.
Discussion of the implementations and the applications is provided in Sec.~\ref{Sec:discussion}, with further examples of reduced shortcuts.
The conclusions are summarized in Sec.~\ref{Sec:conclusions}.

\section{Shortcuts in STIRAP: background}\label{Sec:STIRAP}

\subsection{Shortcuts: General}\label{Sec:General}

\def\c{\mathbf{c}}
\def\a{\mathbf{a}}

We wish to solve the Schr\"odinger equation,
\be\label{SEq}
i \hbar \dot \c(t) = \H(t) \c(t),
\ee
for a system of $N$ states $\ket{\psi_1}, \ket{\psi_2},\ldots,\ket{\psi_N}$, with probability amplitudes $c_k$: $\ket{\c(t)} = [c_1(t),c_2(t),\ldots,c_N(t)]^T$.
Hereafter the overdot denotes the time derivative.
Let the eigenvalues of $\H(t)$ be denoted by $\lambda_k(t)$ and the corresponding (orthonormalized) eigenvectors by $\ket{\phi_k}$ $(k=1,2,\ldots,N)$.
The latter are also known as the adiabatic states and they form an alternative basis (the adiabatic basis) in the $N$-dimensional Hilbert space, i.e. $\ket{\a(t)} = [a_1(t),a_2(t),\ldots,a_N(t)]^T$.
The matrix formed of the eigenvectors of $\H(t)$, viz.
\be\label{W}
\W(t) = [\ket{\phi_1(t)}, \ket{\phi_2(t)}, \ldots, \ket{\phi_N(t)}],
\ee
diagonalizes the Hamiltonian,
\be\label{diag}
\W(t)^\dagger \H(t) \W(t) = \text{diag} [ \hbar\lambda_1(t), \hbar \lambda_2(t), \ldots, \hbar \lambda_N(t) ].
\ee
The transformation $\ket{\c(t)} = \W(t) \ket{\a(t)}$ casts the Schr\"odinger equation into the form
\be
i \hbar \dot \a(t) = \H_a(t) \a(t),
\ee
where
\be\label{Ha}
\H_a(t) = \W(t)^\dagger \H(t) \W(t) - i\hbar \W(t)^\dagger \dot \W(t)
\ee
is the Hamiltonian in the adiabatic basis.
The first term on the RHS is the diagonal matrix \eqref{diag}, and the second term is a matrix comprising the nonadiabatic couplings, i.e. the couplings between the adiabatic states $-i\hbar \langle \phi_k(t) \vert \dot\phi_n(t) \rangle $.
Adiabatic evolution occurs when the system remains in the same adiabatic state in which it is initially.
 The condition for this is that all nonadiabatic couplings linked to this adiabatic state are negligibly small compared to the difference between its eigenvalue and any other eigenvalue, viz.
\be
\left|-i \langle \phi_k(t) \vert \dot\phi_n(t) \rangle \right| \ll \left|\lambda_k(t) - \lambda_n(t) \right|.
\ee
Nonzero nonadiabatic couplings lead to population leaks (nonadiabatic losses) from the populated adiabatic state and ensuing loss of transfer efficiency.

In the ``shortcut to adiabaticity'' concept, an additional term $\H_s(t)$ is added to the original Hamiltonian in Eq.~\eqref{SEq} to obtain a new Hamiltonian $\H'(t) = \H(t) + \H_s(t)$.
The shortcut term $\H_s(t)$ is chosen such that in the basis of the eigenstates $\ket{\phi_k(t)}$ $(k=1,2,\ldots,N)$ of the \emph{original} Hamiltonian $\H(t)$ the nonadiabatic couplings $-i\hbar \langle \phi_k(t) \vert \dot\phi_n(t) \rangle $ are canceled by the additional terms coming from the shortcut $\H_s(t)$.
Specifically, by replacing $\H(t)$ by $\H'(t)$ in Eq.~\eqref{Ha} we find
\be\label{H-full}
\H'_a(t) = \W(t)^\dagger [\H(t) + \H_s(t)] \W(t) - i\hbar \W(t)^\dagger \dot\W(t).
\ee
The ``shortcuts to adiabaticity'' approach imposes the condition
\be\label{H-sta}
\W(t)^\dagger \H_s(t) \W(t) = i\hbar \W(t)^\dagger \dot\W(t).
\ee
Therefore the shortcut reads
\be\label{H-s}
\H_s(t) = i\hbar [\dot\W(t)] \W(t)^\dagger,
\ee
and it leads to the diagonal matrix
\be
\H'_a(t) = \W(t)^\dagger \H(t) \W(t) ,
\ee
see Eq.~\eqref{diag}.

It is important to note that this cancellation happens in the adiabatic basis of the \emph{original} Hamiltonian $\H(t)$, and \emph{not} in the adiabatic basis of the new Hamiltonian $\H'(t)$.
Therefore, the resulting evolution, generated by the new Hamiltonian $\H'(t)$ is \emph{nonadiabatic}.
%To this end, the term ``shortcut to adiabaticity'' is incorrect.
%
Nonetheless, this approach produces a quantum control method for complete population transfer, which can be useful in certain situations.

In general, the shortcut term $\H_s(t)$ can give contributions to all elements of the new Hamiltonian $\H'(t)$, % in the diabatic and adiabatic bases,
 thereby creating a rather messy picture.
Shortcut couplings between various states may be difficult, or even impossible, to implement.
In some special cases, to be considered here, a smart choice of $\H_s(t)$ --- different from the prescription of Eq.~\eqref{H-s} --- can lead to feasible physical implementations, still maintaining very high efficiency of the process.
 %simple analytic formulas for its elements.

\subsection{Shortcut to three-state STIRAP}\label{Sec:STIRAP-3}

The standard STIRAP process operates in a resonant three-state chainwise-connected system, for which the Hamiltonian reads
\be\label{H3}
\H = \tfrac12 \hbar \left[\begin{array}{ccc}
 0 & \Omega _P & 0 \\
 \Omega _P & 0 & \Omega _S \\
 0 & \Omega _S & 0
\end{array}\right],
\ee
where $\Omega_P$ is the (pump) Rabi frequency of the coupling between states 1 and 2, and $\Omega_S$ is the (Stokes) Rabi frequency for the transition $2\fromto 3$, see Fig.~\ref{fig:3ss}.
Both $\Omega_P(t)$ and $\Omega_S(t)$ are assumed real and positive.
The system is initially in state 1 and the objective is to transfer the population to state 3.
The eigenvalues of the Hamiltonian \eqref{H3} are $\lambda_0 = 0$, $\lambda_\pm = \pm \Lambda(t)/2$,
 where $\Lambda(t)=\sqrt{\Omega_P(t)^2 + \Omega_S(t)^2}$ is the rms Rabi frequency.
In terms of the mixing angle $\theta$ defined by
\be\label{theta}
\theta(t) = \arctan \frac{\Omega_P(t)}{\Omega_S(t)},
\ee
the eigenvectors of the Hamiltonian \eqref{H3} read
\bse
\begin{align}
\ket{\phi_0(t)} &= [\cos\theta(t), 0, -\sin\theta(t)]^T, \\
\ket{\phi_+(t)} &= [\sin\theta(t), 1, \cos\theta(t)]^T/\sqrt{2}, \\
\ket{\phi_-(t)} &= [\sin\theta(t), -1, \cos\theta(t)]^T/\sqrt{2}.
\end{align}
\ese
For counterintuitively ordered pulses --- Stokes before pump --- the mixing angle $\theta(t)$ changes from 0 initially to $\pi/2$ in the end.
Correspondingly, the zero-eigenvalue eigenstate $\ket{\phi_0(t)}$ changes from $[1,0,0]^T = \psi_1$ initially to $[0,0,-1]^T = -\psi_3$ in the end, thereby providing an adiabatic connection between states 1 and 3.
If the evolution is adiabatic, then the system will remain in the adiabatic state $\ket{\phi_0(t)}$ at all times and the population will pass from state 1 to state 3.
An added bonus of STIRAP is that state $\ket{\phi_0}$ has no component of the middle state 2, which is usually a lossy state; hence the name ``dark'' state for $\ket{\phi_0(t)}$.
Therefore, in the adiabatic limit no population loss occurs during the population transfer process if states 1 and 3 are ground or metastable, as they usually are.

In the general nonadiabatic regime, there exist nonadiabatic couplings $\pm i \dot\theta/\sqrt{2}$ between the dark state and the other two adiabatic states which generate population leaks from the dark state $\ket{\phi_0(t)}$ with an ensuing loss of population transfer efficiency.
In order to suppress them, one demands the (local) adiabatic condition $\Lambda(t) \gg \dot\theta(t)$.
By integrating over time, one finds the (global) adiabatic condition $A \gg \pi$, where $A = \int_{-\infty}^\infty  \Lambda(t) dt$ is the rms pulse area.
Hence, large pulse areas are needed for high population transfer efficiency.

To this end, Unanyan \etal~\cite{Unanyan1997} proposed to add a Q field, which shortcuts the transition $1\fromto 3$, as shown in Fig.~\ref{fig:3ss}.
% If this field is real, or has an arbitrary phase, the picture in the basis $\{\ket{\phi_-(t)},\ket{\phi_0(t)},\ket{\phi_+(t)}\}$ becomes rather messy.
The Q field has a phase shift of $\pi/2$ relative to the P and S fields. % (assumed real and positive)
The Hamiltonian becomes
\be\label{H3Q}
\H' = \tfrac12 \hbar \left[\begin{array}{ccc}
 0 & \Omega _P & i \Omega_Q \\
 \Omega _P & 0 & \Omega _S \\
 -i \Omega_Q & \Omega _S & 0
\end{array}\right].
\ee
Furthermore, if \cite{Unanyan1997}
\be\label{STIRAP-shortcut}
\Omega_Q(t) = 2\dot\theta(t),
\ee
then the nonadiabatic coupling is completely cancelled by the Q field.
Then the system will stay in the dark state $\ket{\phi_0(t)}$ and the population transfer $1\to 3$ will take place with unit probability --- this is an exact rather than approximate result.
It is easy to verify that Eqs.~\eqref{H3Q} and \eqref{STIRAP-shortcut} are exactly what the general procedure of Eq.~\eqref{H-s} prescribes.
Figure \ref{fig:3ss}(bottom) illustrates the shortcut pulse for Gaussian-shaped P and S pulses.

In the next two sections I extend the three-state shortcut-STIRAP to STIRAP-like process in five-state chainwise-connected systems of practical significance.
%: when the five states are the sublevels $m_g=0,\pm 2$ of a ground level with an angular momentum $J_g=2$ and the sublevels $m_e=\pm 1$ of an excited level with $J_e=1$ or $J_e=2$.
%Because the electric-dipole transition moments are proportional to different Clebsch-Gordan coefficients in these two cases, the shortcuts differ too.
% Further examples for shortcut-STIRAP in five-state systems are presented in Sec.~\ref{Sec:discussion}.

%\section{Shortcut-STIRAP in $J_g=2 \fromto J_e=1$ system}\label{Sec:STIRAP-2-1}

\section{Multistate STIRAP: Standard ``shortcuts to adiabaticity''}\label{Sec:standard}

\subsection{System}

%***************************************************************
\begin{figure}[t]
\includegraphics[width=0.60\columnwidth]{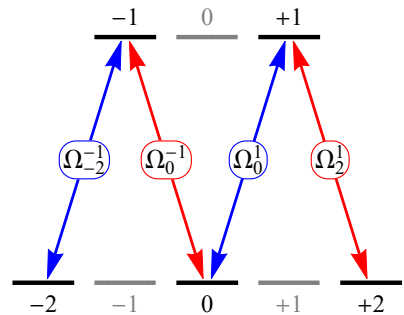}
\caption{
M-shaped five-state chainwise-connected system formed by the magnetic sublevels $m_g=\pm 2,0$ of a ground level with an angular momentum $J_g=2$ and $m_e=\pm 1$ of an excited level with an angular momentum $J_e = 1$.
The transitions $m_g=-2 \fromto m_e=-1$ and $m_g=0 \fromto m_e=+1$ are driven by a $\sigma^+$ polarized laser field, while
the transitions $m_g=0 \fromto m_e=-1$ and $m_g=+2 \fromto m_e=+1$ are driven by a $\sigma^-$ polarized laser field.
}
\label{fig:scheme-21}
\end{figure}
%***************************************************************

Consider the five-state chainwise-connected system formed of the sublevels $m_g=0,\pm 2$ of a ground level with an angular momentum $J_g=2$ and the sublevels $m_e=\pm 1$ of an excited level with $J_e=1$ driven by two left ($\sigma^-$) and right ($\sigma^+$) circularly polarized laser fields, see Fig.~\ref{fig:scheme-21}.
The Hamiltonian driving this system reads
\be\label{H21}
\H = \tfrac12 \hbar \left[\begin{array}{ccccc}
 0 & \Omega _{-2}^{-1} & 0 & 0 & 0 \\
 \Omega _{-2}^{-1} & 0 & \Omega _{0}^{-1} & 0 & 0 \\
 0 & \Omega _{0}^{-1} & 0 & \Omega _{0}^{1} & 0 \\
 0 & 0 & \Omega _{0}^{1} & 0 & \Omega _{2}^{1} \\
 0 & 0 & 0 & \Omega _{2}^{1} & 0
\end{array}\right],
\ee
where $\Omega_{m_g}^{m_e}$ is the Rabi frequency of the coupling between sublevels $m_g$ and $m_e$.
The transitions $m_g=-2 \fromto m_e=-1$ and $m_g=0 \fromto m_e=1$ are driven by the $\sigma^+$ field, while
the transitions $m_g=0 \fromto m_e=-1$ and $m_g=2 \fromto m_e=1$ are driven by the $\sigma^-$ field, see Fig.~\ref{fig:scheme-21}.
The Rabi frequencies are proportional to the respective Clebsch-Gordan coefficients, $\Omega _{-2}^{-1} = \xi_{-2}^{-1}\,\Omega_P$, $\Omega _{0}^{-1} = \xi_{0}^{-1}\,\Omega_S$, $\Omega _{0}^{1} = \xi_{0}^{1}\,\Omega_P$, $\Omega _{2}^{1} = \xi_{2}^{1}\,\Omega_S$, where $\Omega_P$ and $\Omega_S$ are the Rabi frequency ``units'' associated with the $\sigma^+$ and $\sigma^-$ fields, respectively.
For the $J_g=2 \fromto J_e=1$ system, we have $\xi_{-2}^{-1} = \sqrt{\frac35}$, $\xi_{0}^{-1} = \sqrt{\frac1{10}}$, $\xi_{0}^{1} = \sqrt{\frac1{10}}$, $\xi_{2}^{1} = \sqrt{\frac35}$.
For the $J_g=2 \fromto J_e=2$ system, the linkage pattern is the same but the Clebsch-Gordan coefficients are different: $\xi_{-2}^{-1} = -\sqrt{\frac13}$, $\xi_{0}^{-1} = \sqrt{\frac12}$, $\xi_{0}^{1} = -\sqrt{\frac12}$, $\xi_{2}^{1} = \sqrt{\frac13}$.

The eigenvalues of the Hamiltonian \eqref{H21} read
\bse\label{EV21}
\begin{align}
\lambda_0 &= 0, \\
\lambda_{--} &= -\frac {\Lambda \sqrt{7-r}} {4 \sqrt{5}},\quad \lambda_{+-} = \frac {\Lambda \sqrt{7-r}} {4 \sqrt{5}},\\
\lambda_{++} &= \frac {\Lambda \sqrt{7+r}} {4 \sqrt{5}}, \quad \lambda_{-+} = -\frac {\Lambda \sqrt{7+r}} {4 \sqrt{5}},
\end{align}
for $J_g=2 \fromto J_e=1$, and
\ese
\bse\label{EV22}
\begin{align}
\lambda_0 &= 0, \\
\lambda_{--} &= -\frac {\Lambda \sqrt{5-s}} {4 \sqrt{3}},\quad \lambda_{+-} = \frac {\Lambda \sqrt{5-s}} {4 \sqrt{3}},\\
\lambda_{++} &= \frac {\Lambda \sqrt{5+s}} {4 \sqrt{3}}, \quad \lambda_{-+} = -\frac {\Lambda \sqrt{5+s}} {4 \sqrt{3}},
\end{align}
\ese
for $J_g=2 \fromto J_e=2$.
Here $\Lambda = \sqrt{\Omega_P^2 + \Omega_S^2}$, $r=\sqrt{13+12 \cos 4\theta}$, and $s=\sqrt{5-4 \cos 4\theta}$.
The mixing angle $\theta$ is introduced by Eq.~\eqref{theta}.
Note that $0 \leqq \theta(t) \leqq \pi/2$.
Obviously, the relations $\lambda_{-+} < \lambda_{--} <\lambda_0 < \lambda_{+-} < \lambda_{++}$ apply in both cases due to $r>0$ and $s>0$.

The eigenstates $\ket{\phi_{xy}(t)}$ of the Hamiltonian \eqref{H21} corresponding to the eigenvalues $\lambda_{xy}$ of Eqs.~\eqref{EV21} or \eqref{EV22} with $x,y=\pm$, are too cumbersome to be presented here but they are straightforward to calculate.
The (normalized) dark state reads
\be\label{dark-21}
\ket{\phi_0} = \frac {\left[\sqrt{2} \cos ^2\theta,0,-\sqrt{3} \sin 2 \theta, 0, \sqrt{2} \sin ^2\theta \right]^T} {\sqrt{3- \cos 4\theta }}
\ee
for $J_g=2 \fromto J_e=1$ and
\be\label{dark-22}
\ket{\phi_0} = \frac {\left[\sqrt{6} \cos ^2 \theta ,0, \sin 2 \theta ,0,\sqrt{6} \sin ^2 \theta \right]^T} {\sqrt{5+ \cos 4 \theta }}
\ee
for $J_g=2 \fromto J_e=2$.
If the S ($\sigma^-$) pulse precedes the P ($\sigma^+$) pulse, the dark state in both cases will be equal to state $\ket{m_g=-2}$ initially ($\theta=0$) and state $\ket{m_g=2}$ in the end ($\theta=\pi/2$), thereby providing the adiabatic path for complete population transfer from $\ket{m_g=-2}$ to $\ket{m_g=2}$ in the adiabatic limit.

The nonadiabatic couplings $\NAC_{xy} = -i\hbar \langle \phi_{xy}(t) \vert \dot\phi_0(t) \rangle $ between the dark state and the other four eigenstates of the Hamiltonian read
\bse\label{NAC21}
\begin{align}
\NAC_{--} = \NAC_{+-} &= -\frac {i \sqrt{6}\, (1+4 \cos 2 \theta +r) \cos \theta} {\sqrt{(3-\cos 4 \theta) (7-r) (r^2+5r \cos 2 \theta )}} , \\
\NAC_{++} = \NAC_{-+} &= -\frac {i \sqrt{6}\, (1+4 \cos 2 \theta -r) \cos \theta} {\sqrt{(3-\cos 4 \theta) (7+r) (r^2 -5r \cos 2 \theta )}} .
\end{align}
\ese
for $J_g=2 \fromto J_e=1$ and
\bse\label{NAC22}
\begin{align}
\NAC_{--} = \NAC_{+-} &= -\frac {i \sqrt{6}\, (3-4 \cos 2 \theta +s) \cos \theta} {\sqrt{(5+\cos 4 \theta) (5-s) (s^2 - s \cos 2 \theta )}} , \\
\NAC_{++} = \NAC_{-+} &= -\frac {i \sqrt{6}\, (3-4 \cos 2 \theta -s) \cos \theta} {\sqrt{(5+\cos 4 \theta) (5+s) (s^2 + s \cos 2 \theta )}} .
\end{align}
\ese
for $J_g=2 \fromto J_e=2$.

Now the objective is to eliminate these nonadiabatic couplings by adding shortcut fields on the direct transitions between the magnetic sublevels of the same level.

\subsection{Standard prescription for shortcuts (type I)}

\begin{figure}[t]
\includegraphics[width=0.9\columnwidth]{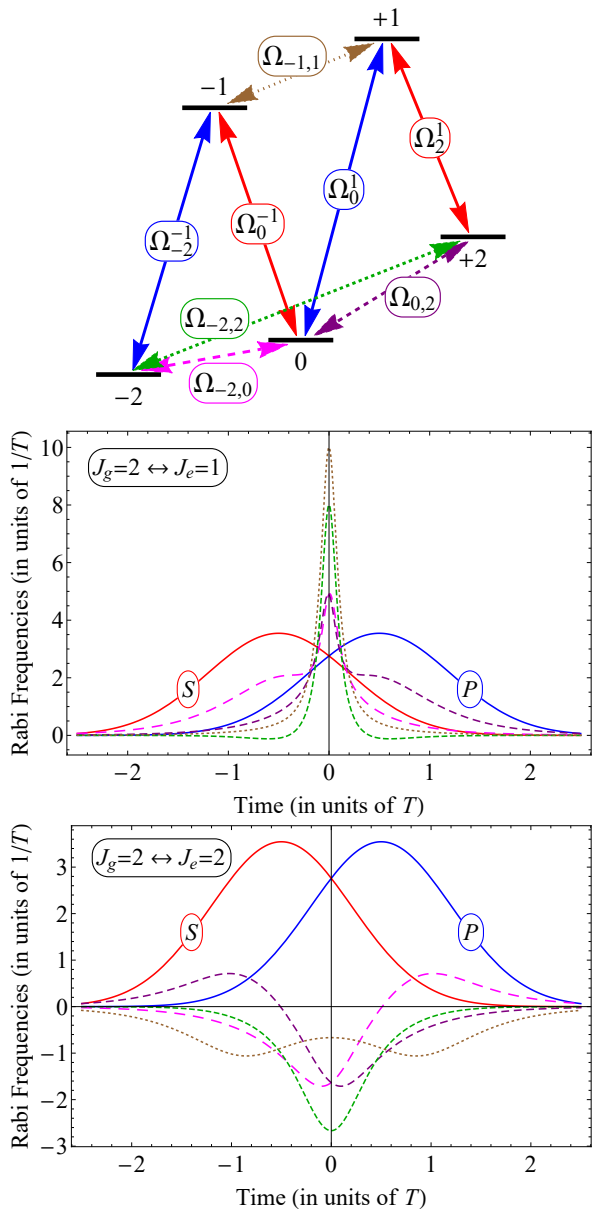}
\caption{
\emph{Top:} Transitions driven by the pump, Stokes and shortcut pulses of \emph{type I}, Eq.~\eqref{H-standard}.
\emph{Middle:} Pulse shapes of the pump (P), Stokes (S) and the four shortcut pulses of Eqs.~\eqref{M21-standard_shortcuts} for Gaussian P and S shapes for the $J_g=2 \fromto J_e=1$ system.
\emph{Bottom:} The same but for the $J_g=2 \fromto J_e=2$ system, Eqs.~\eqref{M22-standard_shortcuts}.
}
\label{fig:shapes-standard}
\end{figure}

The standard prescription of Eq.~\eqref{H-s} leads to the shortcut Hamiltonian
\be\label{H-standard}
\H_s^\prime = \tfrac12 \hbar \left[\begin{array}{ccccc}
 0 & 0 & i\Q_{-2,0} & 0 & i\Q_{-2,2} \\
 0 & 0 & 0 & i\Q^{-1,1} & 0 \\
 -i\Q_{-2,0} & 0 & 0 & 0 & i\Q_{0,2} \\
 0 & -i\Q^{-1,1} & 0 & 0 & 0 \\
 -i\Q_{-2,2} & 0 & -i\Q_{0,2} & 0 & 0
\end{array}\right],
\ee
where
\bse\label{M21-standard_shortcuts}
\begin{align}
\Q_{-2,0} &= \frac{ \sqrt{6}\, (34 + 29 \cos 2 \theta +26 \cos 4 \theta +11 \cos 6 \theta )} {(3- \cos 4 \theta) (13+12 \cos 4 \theta) }  {\dot\theta},\\
\Q_{0,2} &= \frac{ \sqrt{6}\, (34 -29 \cos 2 \theta +26 \cos 4 \theta -11 \cos 6 \theta )} {(3- \cos 4 \theta) (13+12 \cos 4 \theta) }  {\dot\theta},\\
\Q_{-2,2} &= - \frac{4 (1 + 9 \cos 4 \theta ) \sin 2 \theta} {(3- \cos 4 \theta) (13+12 \cos 4 \theta) }  {\dot\theta},\\
\Q^{-1,1} &= \frac{10 }{13 + 12 \cos 4 \theta }  {\dot\theta},
\end{align}
\ese
for $J_g=2 \fromto J_e=1$, and
\bse\label{M22-standard_shortcuts}
\begin{align}
\Q_{-2,0} &= \frac{ \sqrt{6}\, (10 + 9 \cos 2 \theta -14 \cos 4 \theta - \cos 6 \theta )} {(4\cos 4 \theta - 5) (\cos 4 \theta + 5)}  {\dot\theta},\\
\Q_{0,2} &= \frac{ \sqrt{6}\, (10 - 9 \cos 2 \theta -14 \cos 4 \theta + \cos 6 \theta )} {(4\cos 4 \theta - 5) (\cos 4 \theta + 5)}  {\dot\theta},\\
\Q_{-2,2} &= \frac{12 ( 5 - 3 \cos 4 \theta ) \sin 2 \theta} {(4\cos 4 \theta - 5) (\cos 4 \theta + 5)}  {\dot\theta},\\
\Q^{-1,1} &= \frac{6 }{4 \cos 4 \theta - 5}  {\dot\theta},
\end{align}
\ese
for $J_g=2 \fromto J_e=2$.
Therefore, as many as four different shortcut fields are required to satisfy the prescription of Eq.~\eqref{H-s}.
These pulse shapes are displayed in Fig.~\ref{fig:shapes-standard} for Gaussian pump and Stokes pulses,
\bse\label{Gaussian}
\begin{align}
\Omega_P &= \Omega_0 \exp[-(t-\tau/2)^2/T^2], \\
\Omega_S &= \Omega_0 \exp[-(t+\tau/2)^2/T^2],
\end{align}
\ese
 with $\tau=T$.
In the numeric simulations shown in the figures below, the peak amplitude of these pulses is taken as $\Omega_0 = 10\sqrt{\pi}/T$, so that the pulse areas are $A_{P,S} = \int_{-\infty}^{\infty} \Omega_{P,S}(t) dt = 10\pi$.
These values make the evolution nearly (but not perfectly) adiabatic.

%***************************************************************
%***************************************************************

The shortcuts derived above ensure that if the five-state system is initially in \emph{any} adiabatic state $\ket{\phi_k(t)}$ then it will remain in it throughout the evolution.
The price to pay is the necessity of having as many as four additional shortcut fields of Eqs.~\eqref{M21-standard_shortcuts} or \eqref{M22-standard_shortcuts}, which is a rather large increase compared to the single shortcut field needed in three-state STIRAP, Fig.~\ref{fig:3ss}.
Actually, this is an overkill for the problem posed here --- complete population transfer from state $m_g=-2$ to state $m_g=2$ --- because the population transfer proceeds through just a single adiabatic state: the dark state $\ket{\phi_0(t)}$.
In order to achieve this objective, it is sufficient to cancel only the nonadiabatic couplings $\NAC_{xy} = -i\hbar \langle \phi_{xy}(t) \vert \dot\phi_0(t) \rangle $, with $x,y=\pm$, related to the dark state $\ket{\phi_0(t)}$, see Eqs.~\eqref{NAC21} or \eqref{NAC22}.
This approach, which leads to fewer shortcut fields, is considered below.

\section{Multistate STIRAP: Reduced shortcuts}\label{Sec:reduced}

\subsection{Derivation of reduced shortcuts}

We start from Eq.~\eqref{H-sta} the fulfillment of which ensures the cancellation of all nonadiabatic couplings, contained in the matrix on the right-hand side of this equation.
Following the arguments above, we wish to cancel only the nonadiabatic couplings connected to the dark state $\ket{\phi_0(t)}$.
By recalling the composition of the transformation matrix $\W(t)$ in Eq.~\eqref{W} we take in Eq.~\eqref{H-sta} only the row of $\W(t)^\dagger$ composed of $\bra{\phi_0(t)}$ to find
\be
\bra{\phi_0 (t)} \H_s(t) \W(t) = i\hbar \bra{\phi_0 (t)} \dot\W(t),
\ee
and hence
\be
\bra{\phi_0 (t)} \H_s(t) = i\hbar \bra{\phi_0 (t)} \dot\W(t) \W(t)^\dagger.
\ee
After Hermitian conjugation we obtain
\be\label{H-reduced}
\H_s(t)^\dagger \ket{\phi_0 (t)} = -i\hbar \W(t) \dot\W(t)^\dagger \ket{\phi_0}.
\ee
This equation represents a set of linear algebraic equations from which we can find $\H_s(t)$.
I present below two solutions to Eq.~\eqref{H-reduced} which involve just two shortcut fields, rather than four as in the standard prescription of Eqs.~\eqref{M21-standard_shortcuts}.

%Following the standard prescription of 4 fields of Eqs.~\eqref{M21-standard_shortcuts}, I

\subsection{Shortcuts of type II}

%***************************************************************
\begin{figure}[t]
\includegraphics[width=0.9\columnwidth]{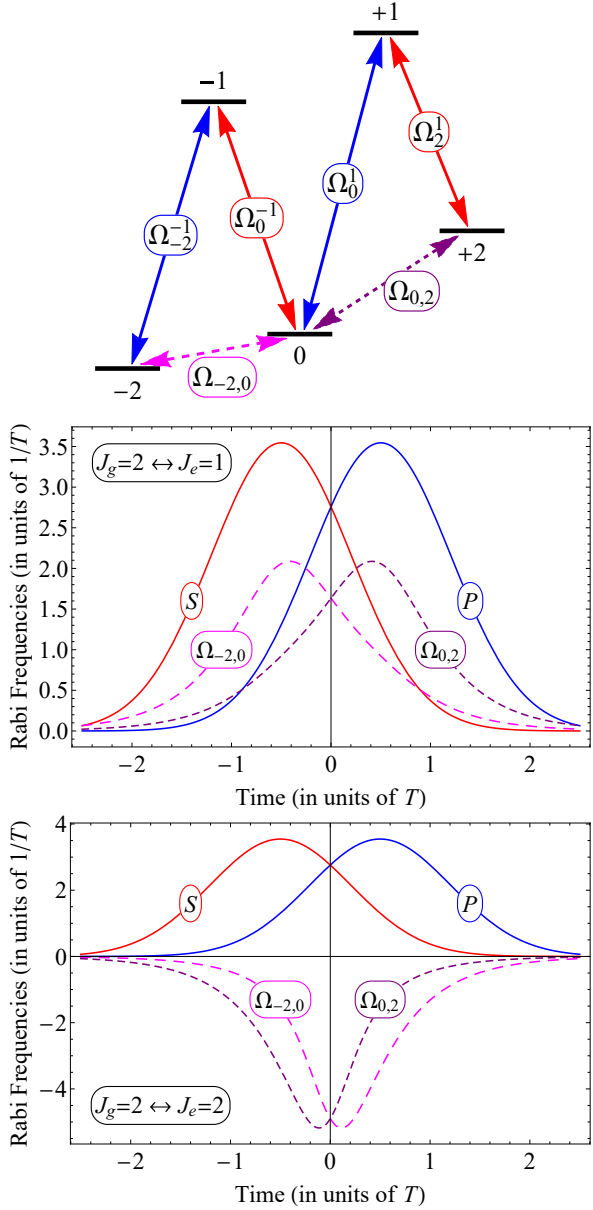}
\caption{
\emph{Top:} Transitions driven by the pump, Stokes and shortcut pulses of \emph{type II}, Eq.~\eqref{H-13-35}.
\emph{Middle:} Pulse shapes of the pump (P), Stokes (S) and the two shortcut pulses of Eqs.~\eqref{M21-shortcuts} for Gaussian P and S shapes for the $J_g=2 \fromto J_e=1$ system.
\emph{Bottom:} The same but for the $J_g=2 \fromto J_e=2$ system, Eqs.~\eqref{M22-shortcuts}.
}
\label{fig:shapes-13-35}
\end{figure}
%***************************************************************

\begin{figure}[t]
\includegraphics[width=0.85\columnwidth]{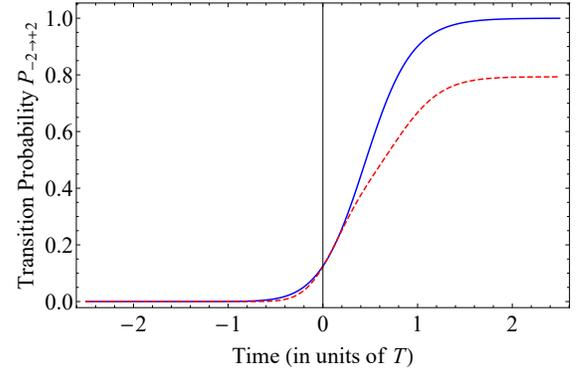}
\caption{
Time evolution of the transition probability $P_{-2 \to 2}$ for the $J_g=2 \fromto J_e=1$ system.
Solid curve: with the shortcuts \eqref{M21-shortcuts} as in Fig.~\ref{fig:shapes-13-35}.
Dashed curve: no shortcuts.
The P and S pulses are Gaussian, Eq.~\eqref{Gaussian}.
%, with pulse areas of $A_{P,S} = \int_{-\infty}^{\infty} \Omega_{P,S}(t) dt = 10\pi$.
}
\label{fig:time}
\end{figure}

Consider two independent shortcut couplings between the adjacent dark-state sublevels produced by two shortcut fields (\emph{type II} shortcuts).
For our system, consider the two shortcut couplings $\Q_{-2,0}$ on the transitions $m_g=-2 \fromto m_g=0$ and $\Q_{0,2}$ on the transition $m_g=0 \fromto m_g=2$, as shown in Fig.~\ref{fig:shapes-13-35}(top).
They give rise to the Hamiltonian
\be\label{H-13-35}
\H_s^{\prime\prime} = \tfrac12 \hbar \left[\begin{array}{ccccc}
 0 & 0 & i\Q_{-2,0} & 0 & 0 \\
 0 & 0 & 0 & 0 & 0 \\
 -i\Q_{-2,0} & 0 & 0 & 0 & i\Q_{0,2} \\
 0 & 0 & 0 & 0 & 0 \\
 0 & 0 & -i\Q_{0,2} & 0 & 0
\end{array}\right].
\ee
The solution of Eq.~\eqref{H-reduced} for $\Q_{-2,0}$ and $\Q_{0,2}$ reads
\bse\label{M21-shortcuts}
\begin{align}
\Q_{-2,0} &= 4 \sqrt{\frac23}\, \frac{ 2 + \cos 2\theta }{3 - \cos 4\theta}\, \dot\theta,\\
\Q_{0,2} &= 4 \sqrt{\frac23}\, \frac{ 2 - \cos 2\theta }{3 - \cos 4\theta}\, \dot\theta,
\end{align}
\ese
for $J_g=2 \fromto J_e=1$, and
\bse\label{M22-shortcuts}
\begin{align}
\Q_{-2,0} &= -4 \sqrt{6}\, \frac{ 2 - \cos 2\theta }{5 + \cos 4\theta}\, \dot\theta,\\
\Q_{0,2} &= -4 \sqrt{6}\, \frac{ 2 + \cos 2\theta }{5 + \cos 4\theta}\, \dot\theta,
\end{align}
\ese
for $J_g=2 \fromto J_e=2$.
For Gaussian P and S pulse shapes, the shortcut pulses are shown in Fig.~\ref{fig:shapes-13-35}.
The time evolution of the transition probability $P_{-2 \to 2}$ for the $J_g=2 \fromto J_e=1$ system is shown in Fig.~\ref{fig:time}.
Without the shortcuts the evolution is not adiabatic and the transition probability $P_{-2 \to 2}$ reaches only about 80\%.
With the shortcut fields the transfer efficiency reaches 100\%.
Note that a possible coupling generated by the shortcut fields on the upper transition $m_e=-1 \fromto m_e=+1$ has no effect because the dark state does not contain these sublevels.

%***************************************************************
%***************************************************************

\subsection{Shortcuts of type III}

%***************************************************************
\begin{figure}[t]
\includegraphics[width=0.9\columnwidth]{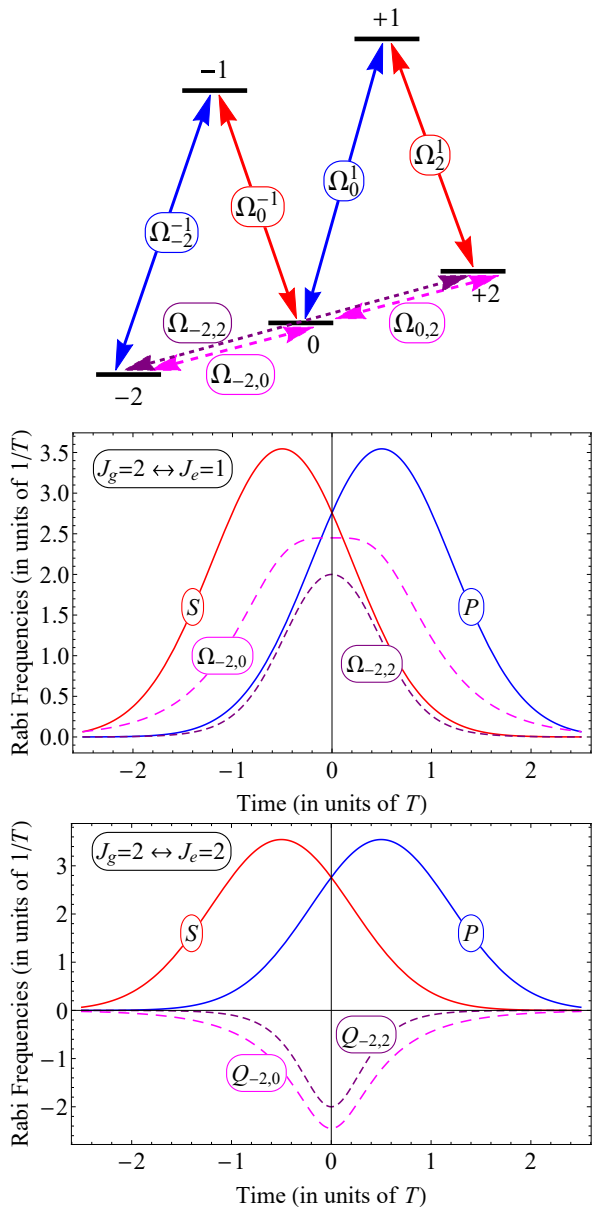}
\caption{
\emph{Top:} Transitions driven by the pump, Stokes and shortcut pulses of \emph{type III}, Eq.~\eqref{H-15}.
\emph{Middle:} Pulse shapes of the pump (P), Stokes (S) and the two shortcut pulses of Eqs.~\eqref{M21-shortcuts-15} for Gaussian P and S shapes for the $J_g=2 \fromto J_e=1$ system.
\emph{Bottom:} The same but for the $J_g=2 \fromto J_e=2$ system, Eqs.~\eqref{M22-shortcuts-15}.
}
\label{fig:shapes-15}
\end{figure}
%***************************************************************

Let us now assume that there are three shortcut couplings between the dark-state sublevels produced by two shortcut fields: one field generates the two couplings between the adjacent sublevels and another field generates the coupling between the two end sublevels (\emph{type III} shortcuts).
In our system, let us assume that two shortcut couplings are equal, $\Q_{-2,0}(t) =\Q_{0,2}(t)$, and the other shortcut $\Q_{-2,2}(t)$, which couples $m_g=-2$ and $m_g=+2$ directly, is independent, see Fig.~\ref{fig:shapes-15}(top).
The shortcut Hamiltonian reads
\be\label{H-15}
\H_s = \tfrac12 \hbar \left[\begin{array}{ccccc}
 0 & 0 & i\Q_{-2,0} & 0 & i\Q_{-2,2} \\
 0 & 0 & 0 & 0 & 0 \\
 -i\Q_{-2,0} & 0 & 0 & 0 & i\Q_{0,2} \\
 0 & 0 & 0 & 0 & 0 \\
 -i\Q_{-2,2} & 0 & -i\Q_{0,2} & 0 & 0
\end{array}\right].
\ee
The solution of Eq.~\eqref{H-reduced} for $\Q_{-2,0}$ and $\Q_{-2,2}$ reads
\bse\label{M21-shortcuts-15}
\begin{align}
\Q_{-2,0} =\Q_{0,2} &= \frac{ 4 \sqrt{6} }{3 - \cos 4\theta}\, \dot\theta ,\\
\Q_{-2,2} &= \frac{ 8 \sin 2\theta }{3 - \cos 4\theta}\, \dot\theta .
\end{align}
\ese
for $J_g=2 \fromto J_e=1$, and
\bse\label{M22-shortcuts-15}
\begin{align}
\Q_{-2,0} = \Q_{0,2} & = -\frac{ 4 \sqrt{6} }{5 + \cos 4\theta}\, \dot\theta ,\\
\Q_{-2,2} &= -\frac{ 8 \sin 2\theta }{5 + \cos 4\theta}\, \dot\theta .
\end{align}
\ese
for $J_g=2 \fromto J_e=2$.
For Gaussian P and S pulse shapes, the shortcut pulses are shown in Fig.~\ref{fig:shapes-15}.

\section{Discussion}\label{Sec:discussion}

\subsection{Robustness to parameter variations}

%***************************************************************
\begin{figure}[t]
\includegraphics[width=0.85\columnwidth]{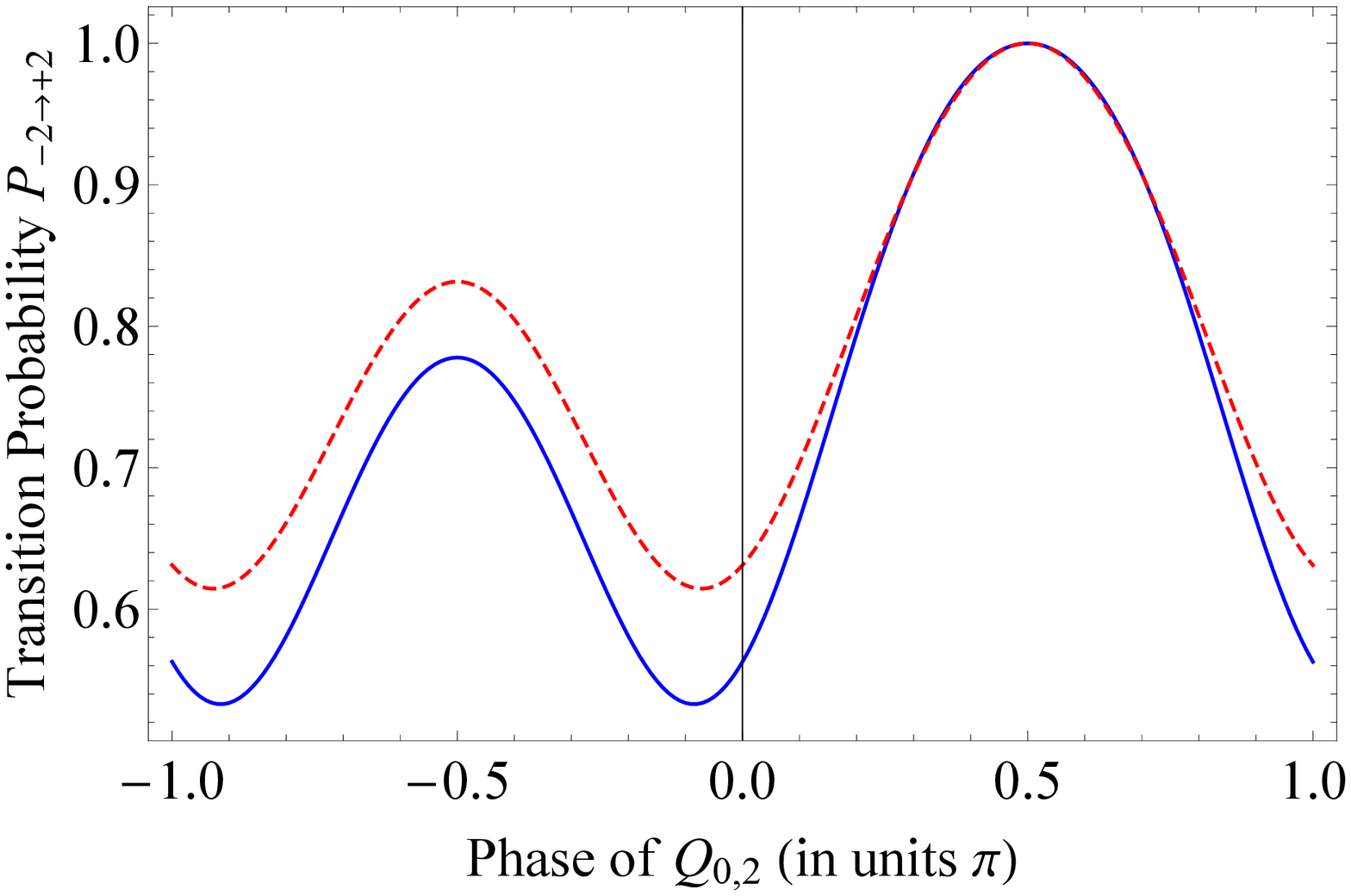}
\caption{
Population transfer efficiency as a function of the phase of the field $\Q_{0,2}$ for type-II shortcuts for $J_g=2 \fromto J_e=1$ system (solid) and $J_g=2 \fromto J_e=2$ system (dashed).
The P and S pulses are Gaussian, Eq.~\eqref{Gaussian}.
}
\label{fig:phase}
\end{figure}
%***************************************************************

It should be pointed out that the shortcut multistate STIRAP technique presented above is strongly dependent on the accurate implementation of the shortcut fields.
Compared to the conventional adiabatic multistate STIRAP the shortcut technique achieves much higher efficiency of population transfer  at the expense of loss of robustness.
Indeed, the shortcut technique is much more sensitive to parameter variations than conventional STIRAP, which is readily found in numerical simulations.

Figure \ref{fig:phase} shows the dependence of the population transfer efficiency $P_{-2 \to 2}$ as a function of the phase of the shortcut field $\Q_{0,2}$ for type-II shortcuts, as in Fig.~\ref{fig:shapes-13-35}.
For a phase of $\pi/2$ the transfer efficiency is 100\% but away from this value it rapidly decreases even below the value of 80\% achieved by standard STIRAP without any shortcuts, see Fig.~\ref{fig:time}.

%***************************************************************
\begin{figure}[t]
\includegraphics[width=0.85\columnwidth]{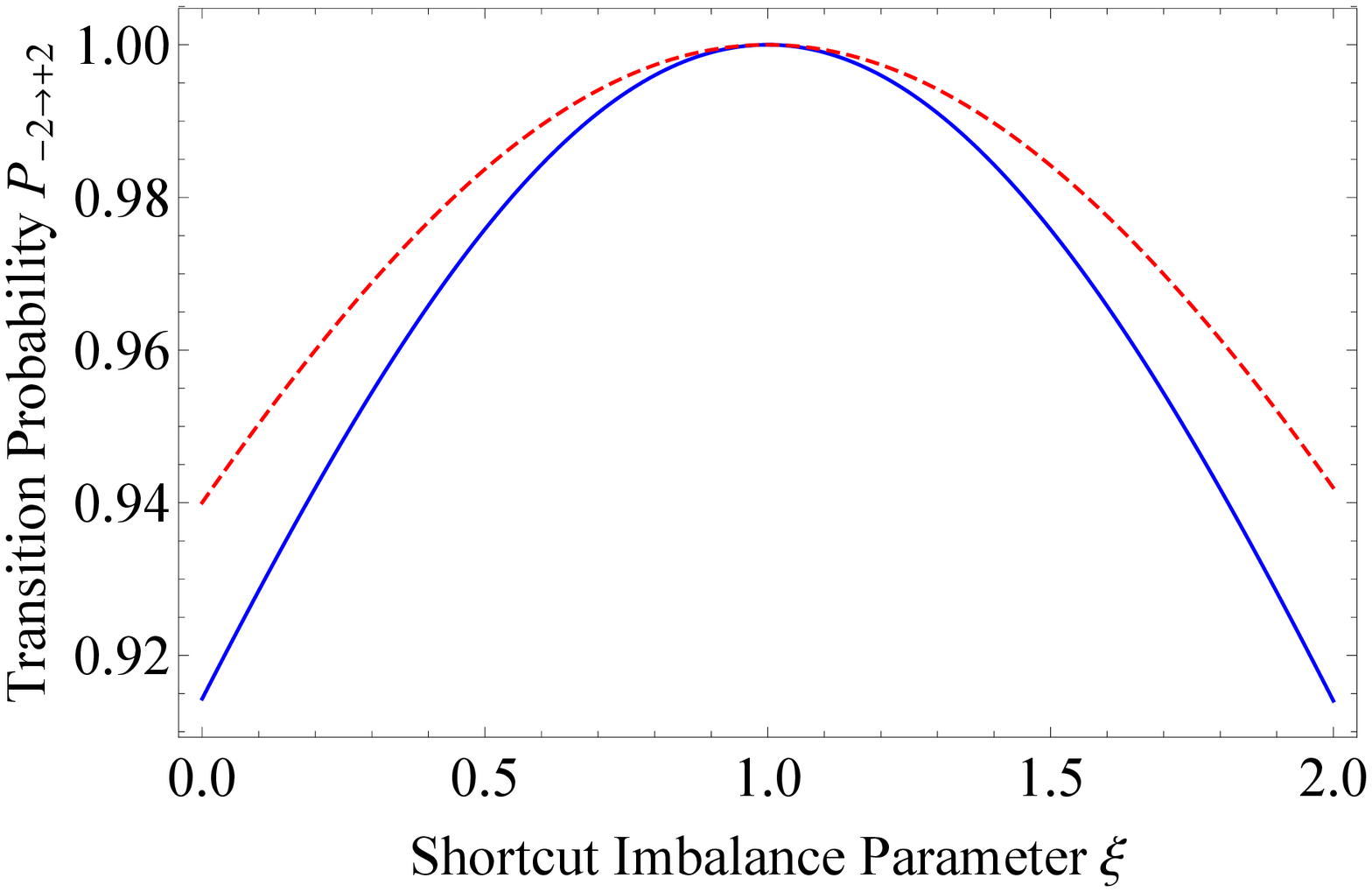}
\caption{
Population transfer efficiency as a function of the amplitude of the field $\Q_{-2,0}$ for type-II shortcuts for $J_g=2 \fromto J_e=1$ system (solid) and $J_g=2 \fromto J_e=2$ system (dashed).
The P and S pulses are Gaussian, Eq.~\eqref{Gaussian}.
}
\label{fig:Q13}
\end{figure}
%***************************************************************

Figure \ref{fig:Q13} shows the dependence of the population transfer efficiency $P_{-2 \to 2}$ on the amplitude of the shortcut field $\Q_{-2,0}$  for type-II shortcuts, as in Fig.~\ref{fig:shapes-13-35}.
Here it is assumed that the shortcut $\Q_{-2,0}$ is replaced by $\xi \Q_{-2,0}$, and $P_{-2 \to 2}$ is plotted versus the imbalance parameter $\xi$.
For $\xi=1$, which is the ideal case, the transfer efficiency is 100\%, but for $\xi \neq 1$ the transfer efficiency decreases.

%***************************************************************
\begin{figure}[t]
\includegraphics[width=0.85\columnwidth]{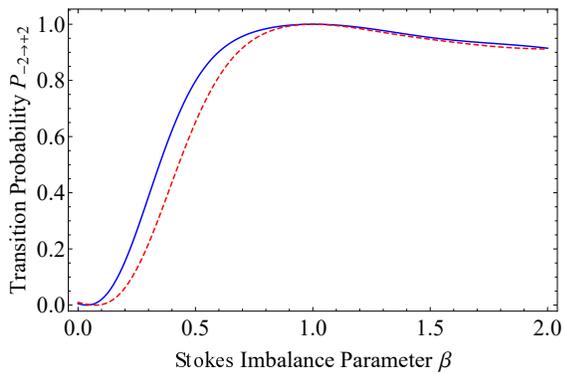}
\caption{
Population transfer efficiency as a function of the amplitude of the Stokes field $S$ for type-II shortcuts for $J_g=2 \fromto J_e=1$ system (solid) and $J_g=2 \fromto J_e=2$ system (dashed).
The P and S pulses are Gaussian, Eq.~\eqref{Gaussian}.
}
\label{fig:Stokes}
\end{figure}
%***************************************************************

Finally, Fig.~\ref{fig:Stokes} shows the population transfer efficiency $P_{-2 \to 2}$ versus the amplitude of the Stokes pulse $S$, again for type-II shortcuts, as in Fig.~\ref{fig:shapes-13-35}.
It is assumed that the Stokes field $S$ is replaced by $\beta S$, and $P_{-2 \to 2}$ is plotted versus the imbalance parameter $\beta$.
For $\xi=1$, which is the balanced case, the transfer efficiency is 100\%, but away from this value the transfer efficiency significantly drops.

\subsection{Feasibility and implementation issues}

Any theoretical proposal should always estimate various implementation issues.
The direct ``shortcut-to-adiabaticity'' approach of type-I shortcuts, Fig.~\ref{fig:shapes-standard} requires four additional very well controlled shortcut fields and it is clearly the most difficult one to implement in a real experiment.
The other two proposals, each requiring two shortcut fields, are obviously the better candidates.
The reduced complexity stems from the fact that only the nonadiabatic couplings related to the dark state are cancelled.
The other nonadiabatic couplings between the other four adiabatic states are irrelevant in the present context of complete population transfer between the two ends of the five-state chain because only the dark state is populated in the ideal case.

The second approach of type-II shortcuts, Fig.~\ref{fig:shapes-13-35}, demands very well controlled shortcut fields on the direct transitions $-2 \fromto 0$ and  $0 \fromto 2$.
These transitions can be driven by rf fields after splitting the magnetic sublevels by a magnetic field.
If the magnetic field is not sufficiently strong, only the first-order Zeeman will show up and the two transitions will have nearly the same transition frequency, thereby ruling out selective driving.
One possibility is to use a stronger magnetic field and then the second-order Zeeman shift will split the two transition frequencies.
The other possibility is to apply an electric field and the combined action of the linear Zeeman effect and the quadratic Stark effect will make the two transition frequencies different again.

The third approach of type-III shortcuts, Fig.~\ref{fig:shapes-15}, demands only a weak magnetic field to split the sublevels $0,\pm 2$ because it assumes that the transitions $-2 \fromto 0$ and  $0 \fromto 2$ are driven by the same field.
The challenge here is generating a well controlled direct coupling between states $-2$ and $2$.
The transition $-2 \fromto 2$ is of much higher order than the $\pm 2 \fromto 0$ transitions and hence much weaker.
However, one can still achieve an effecting coupling $\Q_{-2,2}$ by using two off-resonant fields on the $\pm2 \fromto 0$ transitions.
By adiabatically eliminating state $0$ one obtains an effective coupling for the transition $-2 \fromto 2$.

% \textbf{Couplings of $\pm 1$ sublevels not important}

%***************************************************************
\begin{figure}[t]
\includegraphics[width=0.62\columnwidth]{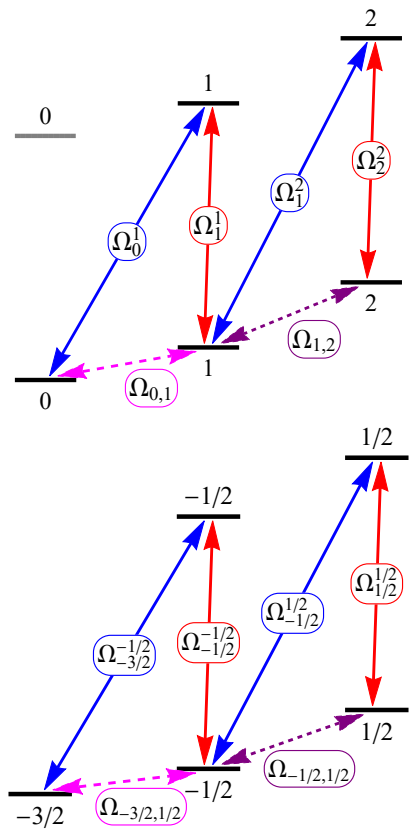}
\caption{
\emph{Top:} Transitions for the $J_g=2 \fromto J_e=2$ system prepared initially in state $\ket{m_g=0}$ and driven by $\sigma^+$ (pump) and $\pi$ (Stokes) polarized laser fields and two shortcut rf fields.
\emph{Bottom:} Transitions for the $J_g=\frac32 \fromto J_e=\frac12$ system prepared initially in state $\ket{m_g=-\frac32}$ and driven by $\sigma^+$ (pump) and $\pi$ (Stokes) polarized laser fields and two shortcut rf fields.
Magnetic sublevels which are not coupled by the driving fields are not shown for the sake of simplicity.
}
\label{fig:sigma-pi}
\end{figure}
%***************************************************************

This work has been mainly concerned with the $\sigma^+\sigma^-$ configuration.
It is straightforward to extend the method to the $\sigma \pi$ configuration too \cite{Featonby1996,Featonby1998,Godun1999,Chalupczak2005}, see Fig.~\ref{fig:sigma-pi}.
For example, $\sigma^+ \pi$ driving of the $J_g = 2 \fromto J_e=2$ system starting in $m_g=0$ will create a five-state chain $m_g=0 \fromto m_e=1 \fromto m_g=1 \fromto m_e=2 \fromto m_g=2$ (owing to the fact that the $m_g=0 \fromto m_e=0$ transition is forbidden, as used in Refs.~\cite{Featonby1996,Featonby1998,Godun1999,Godun1999jpb}).
The Clebsch-Gordan coefficients in this system are $\xi_0^1 = \sqrt{\frac12}$, $\xi_1^1 = \sqrt{\frac16}$, $\xi_1^2 = \sqrt{\frac13}$, $\xi_2^2 = \sqrt{\frac23}$.
The shortcuts are straightforward to calculate.
For example, the ones of type II shown in Fig.~\ref{fig:sigma-pi}(top) are
\bse
\begin{align}
\Q_{0,1} &= \frac{4 \sqrt{3}}{3 - \cos ^4\theta } \dot\theta ,\\
\Q_{1,2} &= \frac{2 \sqrt{2} \left(3 - \cos ^2\theta \right)}{3 - \cos ^4 \theta}  \dot\theta.
\end{align}
\ese

Another example is the  $\sigma^+ \pi$ driving of the $J_g = \frac32 \fromto J_e=\frac12$ system starting in $m_g=-\frac32$ sublevel which will create another five-state chain, see Fig.~\ref{fig:sigma-pi}(bottom).
The Clebsch-Gordan coefficients are $\xi_{-3/2}^{-1/2} = \sqrt{\frac12}$, $\xi_{-1/2}^{-1/2} = -\sqrt{\frac13}$, $\xi_{-1/2}^{1/2} = \sqrt{\frac16}$, $\xi_{1/2}^{1/2} = -\sqrt{\frac13}$.
The shortcuts of type II in Fig.~\ref{fig:sigma-pi}(bottom) read
\bse
\begin{align}
\Q_{-\frac32,-\frac12} &= - \frac{4 \sqrt{6}}{3 + \cos ^4\theta } \dot\theta ,\\
\Q_{-\frac12,\frac12} &= -\frac{2 \sqrt{2} \left(3 + \cos ^2\theta \right)}{3 + \cos ^4 \theta}  \dot\theta.
\end{align}
\ese

This paper has considered specific shortcuts to five-state STIRAP only.
Multistate STIRAP has been demonstrated in nine-state systems of magnetic sublevels in $J_g=4 \fromto J_e=3$ and $J_g=4 \fromto J_e=4$ too \cite{Pillet1993,Valentin1994,Goldner1994,Theuer1998,Featonby1996,Featonby1998,Godun1999,Godun1999jpb}.
The challenge to implement shortcuts in such systems is purely algebraic because the shortcuts cannot be derived in a simple analytic form.
However, numeric derivation of the shortcuts in such systems following the procedures described above should be fairly straightforward.

\section{Conclusions and outlook}\label{Sec:conclusions}

In this paper three types of shortcuts which eliminate the nonadiabatic couplings in multistate STIRAP and enable population transfer with unit efficiency have been derived.
Specifically, two five-state systems formed of the magnetic sublevels of two levels with angular momenta $J_g = 2$ and $J_e = 1$ or $J_e = 2$ driven by two delayed left and right circularly polarized laser pulses have been considered in detail.
In the adiabatic limit, which requires very large pulse areas, multistate STIRAP transfers the population adiabatically between the two end states $M_g=\pm 2$ of the five-state chain via a dark state, as in three-state STIRAP.
For moderately large pulse areas nonadiabatic couplings cause population leaks from the dark state and erode the transfer efficiency.

The application of shortcut fields between the magnetic sublevels belonging to the same level allows one to cancel the nonadiabatic couplings and reach unit transfer efficiency.
Three shortcut choices have been studied here, all of which admit simple analytic solutions for the shortcut fields.
The first one is obtained from the prescription of the ``shortcut-to-adiabaticity'' approach and it prescribes four additional shortcut fields.
The second approach demands two shortcuts for the transitions $m_g=-2\fromto m_g=0$ and $m_g=0\fromto m_g=2$.
The last approach also assumes two different shortcut fields: one acting simultaneously on the transitions $m_g=0\fromto m_g=\pm2$ and another on the transition $m_g=-2\fromto m_g=2$.

All three approaches ensure a unit transfer efficiency $m_g=-2\to m_g=2$ but they have different experimental complexity.
The direct ``shortcut-to-adiabaticity'' approach with its four additional shortcut fields is clearly very demanding, if possible at all in a real experiment.
The second approach requires the application of either a strong magnetic field, so that the second-order Zeeman shift becomes pronounced and allows for the separation of the two transitions $m_g=0\fromto m_g=\pm2$ in the frequency space, or the application of both magnetic and electric fields.
The third approach requires the application of moderate magnetic field only but it comes with the necessity to generate a well controlled coupling between states $m_g=-2$ and $m_g=2$.

The results in this paper can be of interest to applications wherein high efficiency of population transfer is essential.
One such application is quantum information processing \cite{Nielsen2000}.
Another application is STIRAP-based atomic mirrors and beams splitters in atom optics \cite{Lawall1994,Weitz1994prl,Weitz1994pra,Pillet1993,Valentin1994,Goldner1994,Theuer1998,Featonby1996,Featonby1998,Godun1999,Godun1999jpb}.
A promising application of the proposed shortcut multistate STIRAP is in the production of ultracold molecules from ultracold atoms \cite{Winkler2007,Danzl2008,Danzl2010,Ni2008,Lang2008,Aikawa2010,Stellmer2012,Takekoshi2014,Molony2014,Molony2016,Park2015,Guo2016,Rvachov2017,DeMarco2019,Seesselberg2018}
It starts with a mixture of two ultracold atomic species at high phase space density, which are adiabatically associated into a weakly bound Feshbach molecular state.
Then the Feshbach molecules are transferred into the electronic, vibrational, and rotational ground state of the molecules using STIRAP via an intermediate electronically excited state, with a typical efficiency reported hitherto of about 90\%.
Shortcuts can be helpful here both in three-state and multistate STIRAP \cite{Danzl2010} because high transfer efficiency is essential in order to preserve the phase-space density of the ultracold mixture.
Yet another application can be found in the initialization of the clock state $m=0$ in caesium fountain frequency standards operating in the nK temperature range where 97\% efficiency has been reported by multistate STIRAP \cite{Chalupczak2005}.
Cavity QED experiments can also benefit from the high efficiency of shortcut multistate STIRAP, e.g. in quantum-state mapping between multilevel atoms and cavity light fields \cite{Parkins1993,Parkins1995}.

Finally, this paper has focused on five-state systems formed by the magnetic sublevels of angular momentum levels.
The same approach applies to more general systems of arbitrary states if the fields driving the various transitions there can be well controlled.

%%%%%%%%%%%%%%%%%%%%%%%%%%%%%%%%%%%%%%%%%%%%%%%%%%%%%%%%%%%%%%%%%%%%%%%%%%%%%%%%%%%%%%%%%%%%%%%%%%%%%%%%%%%%%%%%%%%%%%%%%%%%%%%%%%%%%%%%%%%%%%%%%%%%%%%%%%%%%%%%
\acknowledgments

This work is supported by the Bulgarian Science Fund Grant No. DN 18/14.
% European Quantum Flagship Project 820314 (MicroQC).
% Bulgarian Science Fund Grant DO02/3 (Quant-ERA Project ERyQSenS).

%\appendix

%%%%%%%%%%%%%%%%%%%%%%%%%%%%%%%%%%%%%%%%%%%%%%%%%%%%%%%%%%%%%%%%%%%%%%%%%%%%%%%%%%%%%%%%%%%%%%%%%%%%%%%%%%%%%%%%%%%%%%%%%%%%%%%%%%%%%%%%%%%%%%%%%%%%%%%%%%%%%%%%
%%%%%%%%%%%%%%%%%%%%%%%%%%%%%%%%%%%%%%%%%%%%%%%%%%%%%%%%%%%%%%%%%%%%%%%%%%%%%%%%%%%%%%%%%%%%%%%%%%%%%%%%%%%%%%%%%%%%%%%%%%%%%%%%%%%%%%%%%%%%%%%%%%%%%%%%%%%%%%%%
%%%%%%%%%%%%%%%%%%%%%%%%%%%%%%%%%%%%%%%%%%%%%%%%%%%%%%%%%%%%%%%%%%%%%%%%%%%%%%%%%%%%%%%%%%%%%%%%%%%%%%%%%%%%%%%%%%%%%%%%%%%%%%%%%%%%%%%%%%%%%%%%%%%%%%%%%%%%%%%%

\end{document}